\documentclass[twocolumn,prb]{revtex4} 
\usepackage{graphicx,float,alltt,psfig,amsmath}

\begin{document}
\newcommand{\be}{\begin{equation}} 
\newcommand{\ee}{\end{equation}}
\newcommand{\bea}{\begin{eqnarray}}
\newcommand{\eea}{\end{eqnarray}}
\newcommand{\om}{\Omega_{\rm M}}
\newcommand{\oll}{\Omega_{\rm \Lambda}}
\newcommand{\omol}{(\Omega_{\rm M}, \Omega_{\rm \Lambda})}
\newcommand{\bctr}{\begin{center}}
\newcommand{\ectr}{\end{center}}
\newcommand{\lsim}{\mbox{$\:\stackrel{<}{_{\sim}}\:$} }
\newcommand{\gsim}{\mbox{$\:\stackrel{>}{_{\sim}}\:$} }

\title{Solutions to the tethered galaxy problem in an expanding
universe and the observation of receding blueshifted objects}
\author{Tamara M. Davis}
\email{tamarad@phys.unsw.edu.au}
\author{Charles H. Lineweaver}
\email{charley@bat.phys.unsw.edu.au}
\author{John K. Webb}
\email{jkw@bat.phys.unsw.edu.au}
\affiliation{Department of Astrophysics, University of New South
Wales, Sydney, 2052, Australia}

\begin{abstract}
We use the dynamics of a galaxy, set up initially at a constant
proper distance from an observer, to derive and illustrate two
counter-intuitive general relativistic results. Although the
galaxy does gradually join the expansion of the universe (Hubble
flow), it does not necessarily recede from us. In particular, in
the currently favored cosmological model, which includes a cosmological 
constant, the galaxy recedes from the observer as it
joins the Hubble flow, but in the previously favored  
cold dark matter model, the galaxy approaches,
passes through the observer, and joins the Hubble flow on the
opposite side of the sky. We show that this behavior is
consistent with the general relativistic idea that space is
expanding and is determined by the acceleration of the expansion
of the universe --- not a force or drag associated with the
expansion itself. We also show that objects at a constant proper
distance will have a nonzero redshift; receding galaxies can be
blueshifted and approaching galaxies can be redshifted.
\vspace{8mm}
\end{abstract}
\vspace{8mm}


\maketitle
\section{INTRODUCTION}
The interpretation of the expansion of the universe in general
relativistic cosmology was, and to some extent still is, the
subject of discussion and controversy. 
Robertson\cite{robertson35} and Walker\cite{walker36} presented
the metric for a homogeneous expanding isotropic universe with a
comoving frame in which receding bodies are at rest, and peculiar
velocities are velocities measured with respect to this comoving
frame. This standard metric and the picture of expanding and
curved space is fully consistent with special relativity locally
and general relativity globally.\cite{harrison93} Milne rejected
the expansion of space and insisted instead on expansion through
space and introduced Newtonian cosmology.\cite{milne34mccrea34}
Although the original formulation was found to be logically
inconsistent,\cite{layzer54} many different formulations of
Newtonian cosmology have since been
proposed.\cite{tipler96tipler96bendean94querella98}  Recession
velocities are a fundamental feature of the general relativistic
expansion of the universe. Harrison\cite{harrison93} has pointed
out a conflict in the use of recession velocities that is resolved
when a distinction is made between the empirical and theoretical
Hubble laws: the empirical redshift distance relation, $cz=HD$, is
valid only at low redshifts, while $v=HD$ derived from the
Robertson-Walker metric is valid for all distances. ($H$ is
Hubble's constant, $v$ is the recession velocity, $z$ is the
redshift, $c$ is the speed of light, and $D$ is the proper
distance.) Perhaps partly because it appears paradoxical and
partly because of the different definitions of distance, recession
velocities greater than $c$ are still a source of much confusion
and skepticism,\cite{mcvittie74weinberg1st3minsschutz85peebles91}
despite several attempts to clarify the
issue.\cite{murdoch77stuckey92kiang97} 

Recently it has been argued that the expansion of space is a
peculiarity of the particular coordinate system used, and the
expansion can equally well be described as an expansion through
space,\cite{page93} or alternately, that the expansion is {\em
locally} kinematical.\cite{peacock99} Debate persists over what
spatial scales participate in the expansion of the
universe,\cite{munley95,shi98tipler99chiueh01dumin02} and the
effect of the expansion of the universe on local systems is a
topic of current
research.\cite{lahav91anderson95cooperstock98hamilton01baker02} 
The general expansion of the universe is known as the Hubble
flow. A persistent confusion is that galaxies set up at rest with
respect to us and then released will start to recede as they pick
up the Hubble flow.  This confusion mirrors the assumption that,
without a force to hold them together, galaxies (and our bodies)
would be stretched as the universe expands. The aim of this paper
is to clarify the nature of the expansion of the universe,
including recession velocities and cosmological redshifts, by
looking at the effect of the expansion on objects that {\em are
not} receding with the Hubble flow. This paper is an extension of
previous discussions on the expansion of
space.\cite{silverman86stuckey92stuckey92bellis93tipler96b,munley95}

To clarify the influence of the expansion of the universe, we
consider the ``tethered galaxy''
problem.\cite{harrison95,peacock01} We set up a distant galaxy at
a constant distance from us and then allow it to move freely. The
essence of the question is, once it has been removed from the
Hubble flow and then let go, what effect, if any, does the
expansion of the Universe have on its movement? In
Sec.~\ref{sect:problem} we derive and illustrate solutions to the
tethered galaxy problem for arbitrary values of the density of the
universe $\om$ and the cosmological constant $\oll$. We show that
no drag is associated with unaccelerated expansion. Our
calculations agree with and generalize the results obtained by
Peacock,\cite{peacock01} but we also point out an interesting
interpretational difference.

The cosmological redshift is important because it is the most
readily observable evidence of the expansion of the universe. In
Sec.~\ref{sect:redshift} we point out a consequence of the fact
that the cosmological redshift is not a special relativistic
Doppler shift, and we derive the counter-intuitive result that a
galaxy at a constant proper distance will have a nonzero
redshift. In Sec.~\ref{sect:obs-cons} we summarize our results and
discuss relativistic radio jets as examples of receding
blueshifted objects.

\begin{figure}
\bctr
\includegraphics[height=76mm,width=86mm]{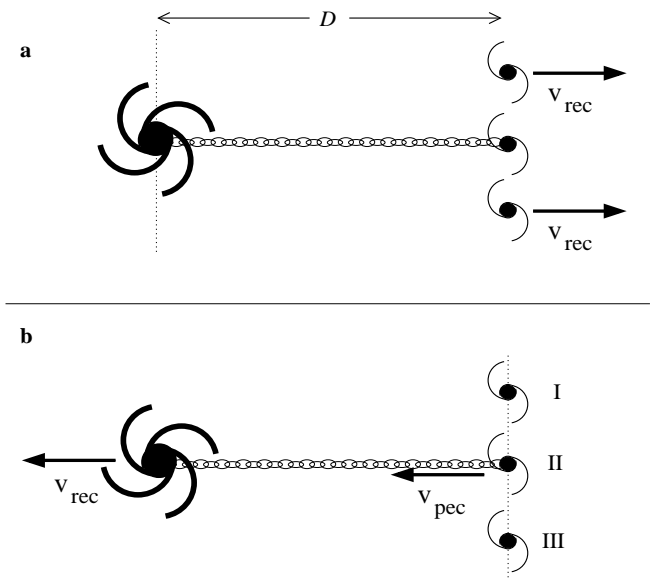}
\caption{\footnotesize{(a) A small distant galaxy (considered to be a massless
test particle) is tethered to an observer in a large galaxy. The
proper distance to the small galaxy, $D$, remains fixed; the small
galaxy does not share the recession velocity of the other galaxies
at the same distance. The tethered galaxy problem is ``What path
does the small galaxy follow when we unhook the tether?'' (b)
Drawn from the perspective of the local comoving frame (out of
which the test galaxy was boosted), the test galaxy has a peculiar
velocity equal to the recession velocity of the large galaxy.
Thus, the tethered galaxy problem can be reduced to ``How far does
an object, with an initial peculiar velocity, travel in an
expanding universe?''}}
\label{fig:problem}
\ectr
\end{figure}

\section{The tethered galaxy problem} \label{sect:problem}
We assume a homogeneous, isotropic universe and use the standard
Friedman-Robertson-Walker (FRW) metric.\cite{peacock99} We only
encounter radial distances, and therefore the FRW metric can be
simplified to
\begin{equation}
ds^2=-c^2dt^2+a^2(t)d\chi^{2}, 
\end{equation}
where $t$ is the proper time of each fundamental observer (also
known as the cosmic time).\cite{weinberg72rindler77} The scale
factor of the universe, $a$, is normalized to 1 at the present
day, $a(t_0)=a_0=1$, and $\chi$ is the comoving coordinate. The
proper distance, $D = a\chi$, is the distance (along a constant
time surface, $dt=0$) between us and a galaxy with comoving
coordinate $\chi$. This is the distance a series of comoving
observers would measure if they each laid their rulers end to end
at the same cosmic instant.\cite{weinberg72rindler77}
Differentiation with respect to proper time is denoted by a dot
and is used to define ``approach'' ($\dot{D}<0$) and ``recede
from'' ($\dot{D}>0$). Present day quantities are given the
subscript zero. Alternative measures of distance are discussed in
Appendix~\ref{ap:distances}.

Figure~\ref{fig:problem} illustrates the tethered galaxy problem.
In an expanding universe distant galaxies recede with recession
velocities given by Hubble's law, $v_{\rm rec} = H D$, where $H$
is the time dependent Hubble constant $H=\dot{a}/a$. We adopt
$H_0=70$\,km\,s$^{-1}$\,Mpc$^{-1}$. Suppose we separate a small
test galaxy from the Hubble flow by tethering it to an observer's
galaxy such that the proper distance between them remains
constant. We neglect all practical considerations of such a tether
because we can think of the tethered galaxy as one that has
received a peculiar velocity boost toward the observer that
exactly matches its recession velocity. We then remove the tether
(or turn off the boosting rocket) to establish the initial
condition of constant proper distance, $\dot{D}_0=0$. The idea of
tethering is incidental, but for simplicity, we refer to this as
the untethered or test galaxy. Note that this is an artificial
setup; we have had to arrange for the galaxy to be moved out of the
Hubble flow in order to apply this zero total velocity condition.
Thus it is not a primordial condition, merely an initial condition
that we have arranged for our experiment. Nevertheless, the
discussion can be generalized to any object that has obtained a
peculiar velocity and in Sec.~\ref{sect:obs-cons} we describe a
similar situation that is found to occur naturally. We define the
total velocity of the untethered galaxy as the time derivative of
the proper distance, $v_{\rm tot} = \dot{D}$, 
\bea
\dot{D} &=& \dot{a}\chi + a\dot{\chi}\\ v_{\rm tot}&=& v_{\rm rec}
+ v_{\rm pec}. \label{eq:vrecvpec}
\eea
The peculiar velocity $v_{\rm pec}$ is the velocity with respect
to the comoving frame out of which the test galaxy was boosted.  It
corresponds to our normal, local notion of velocity and must be
less than the speed of light.  In this section we consider only the nonrelativistic case, $v_{\rm pec}<\hspace{-1.5mm}< c$.  The recession velocity $v_{\rm rec}$
is the velocity of the Hubble flow at the proper distance $D$ and
can be arbitrarily
large.\cite{murdoch77stuckey92kiang97,harrison93} The motion of
this test galaxy reveals the effect the expansion of the universe
has on local dynamics. To enable us to isolate the effect of the
expansion of the universe, we assume that the galaxies have
negligible mass.  By construction the tethered galaxy at an
initial time $t_0$ has zero total velocity, $\dot{D}_0 = 0$, or 
\bea
v_{\rm pec,0} &=& - v_{\rm rec,0}\label{eq:initial1}\\
 a_0\dot{\chi}_0 &=& -\dot{a}_0\chi_0. \label{eq:initial}
\eea
With this initial condition established, we untether the galaxy
and let it coast freely. The question is then: Does the test
galaxy approach, recede, or stay at the same distance?
\begin{figure}
\includegraphics[height=60mm,width=84mm]{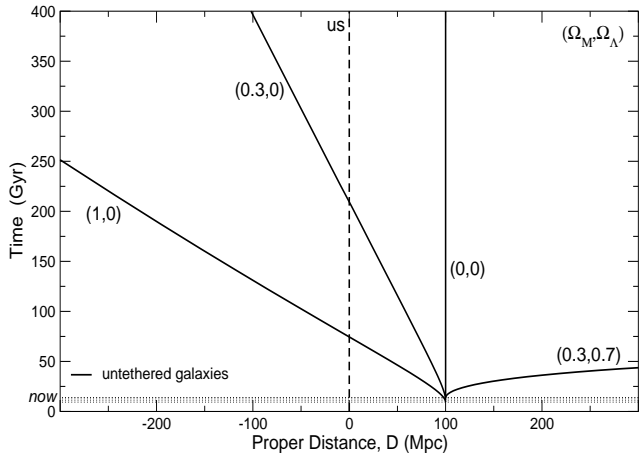}
\caption{\footnotesize{Solutions to the tethered galaxy problem
[Eq.~(\ref{eq:Dsolution})]. For four cosmological models we
untether a galaxy at a distance of $D_0=100$\,Mpc with an initial
peculiar velocity equal to its recession velocity (total initial
velocity is zero) and plot its path. In each case the peculiar
velocity decays as $1/a$. Its final position depends on the model.
In the $\omol=(0.3,0.7)$ accelerating universe, the untethered
galaxy recedes from us as it joins the Hubble flow, while in the
decelerating examples, $\omol=(1,0)\ \mbox{and}\ (0.3,0)$, the
untethered galaxy approaches us, passes through our position and
joins the Hubble flow in the opposite side of the sky. In the
$\omol=(0,0)$ model the galaxy experiences no acceleration and
stays at a constant proper distance as it joins the Hubble flow
[Eq.~(\ref{eq:q})]. In Sec.~\ref{sect:redshift} and
Fig.~\ref{fig:z0vpecvrec} 
we derive and illustrate the
counter-intuitive result that such a galaxy will be blueshifted.
We are the comoving galaxy represented by the thick dashed line
labeled ``us.'' There is a range of values labeled ``now,''
because the current age of the universe is different in each
model.}} \label{fig:Dsolution}
\end{figure}
\begin{figure} \includegraphics[height=60mm,width=84mm]{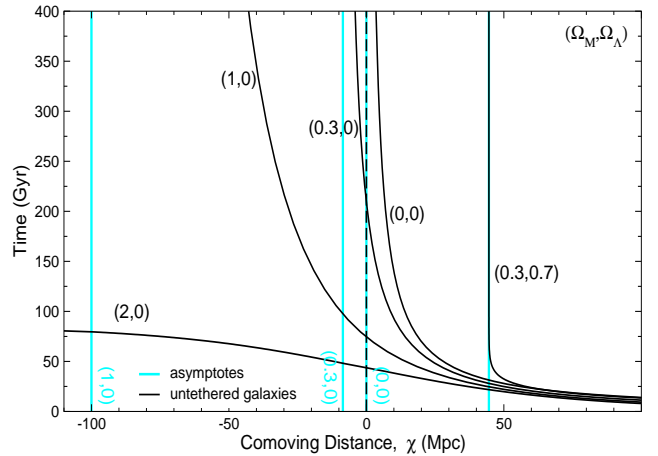}
\caption{\footnotesize{Solutions to the tethered galaxy problem in comoving
coordinates [Eq.~(\ref{eq:chisolution})] for five cosmological
models. In all the models the comoving coordinate of the
untethered galaxy decreases (our initial condition specified a
negative peculiar velocity). In models that do not recollapse the
untethered galaxy coasts and approaches an asymptote as it joins
the Hubble flow. The rate of increase of the scale factor
determines how quickly an object with a peculiar velocity joins
the Hubble flow. In the accelerating universe $\omol = (0.3,0.7)$,
the perturbed galaxy joins the Hubble flow more quickly than in
the decelerating universes $(1,0)$ and $(0.3,0)$, with the $(0,0)$
universe in between. The $\omol = (2,0)$ model is the only model
shown that recollapses. In the recollapsing phase of this model
the galaxy's peculiar velocity increases as $a$ decreases and the
galaxy does not join the Hubble flow [Eq.~(\ref{eq:relativistic})]. In
the (0,0) model the proper distance to the untethered galaxy is
constant, and therefore its comoving distance $\chi=D/a$ tends
toward zero (our position) as $a$ tends toward infinity. The
different models have different starting points in time because
the current age of the universe is different in each model.}}
\label{fig:chisolution}
\end{figure}

The momentum $p$ with respect to the local comoving frame
decays\cite{misner70weinberg72peebles93padmanabhan96} as $1/a$.
This scale factor dependent decrease in momentum is an important
basis for many of the results that follow. For nonrelativistic
velocities $p=mv_{\rm pec}$ (for the relativistic solution see
Appendix~\ref{ap:rel-sol}), and, therefore,
\bea
v_{\rm pec} &=& \frac{v_{\rm pec, 0}}{a} \label{eq:Rinv}\\
a\dot{\chi} &=& \frac{-\dot{a}_0 \chi_0}{a} \label{eq:Rchidot} \\
\chi &=& \;\;\chi_0\biggr[1-\dot{a}_0 \int_{
t_0}^{t}\frac{dt}{a^{2}} \biggr] \label{eq:chisolution}\\ D &=&
a\,\chi_0\biggr[1-\dot{a}_0\int_{ t_0}^{t}\frac{dt}{a^{2}}\biggr]\,.
\label{eq:Dsolution}
\eea
The integral in Eqs.~(\ref{eq:chisolution}) and
(\ref{eq:Dsolution}) can be performed numerically by using
$dt=da/\dot{a}$ and $\dot{a}_0$, where both are obtained directly
from the Friedmann equation,
\begin{equation}
 \dot{a}=\frac{da}{dt}=H_0 \biggr [1+\om(\frac{1}{a}-1)+\oll(a^2-1)\biggr]^{1/2} . \label{eq:fried}
\end{equation}
The normalized matter density $\om=8\pi G\rho_0/3H_0^2$ and the
cosmological constant $\oll=\Lambda/3H_0^2$ are constants
calculated at the present day. The scale factor $a(t)$ is derived
by integrating the Friedmann equation.\cite{felten86}

Equation~(\ref{eq:Dsolution}) provides the general solution to the
tethered galaxy problem. Figure~\ref{fig:Dsolution} shows this
solution for four different models. In the currently favored
model, $\omol=(0.3,0.7)$, the untethered galaxy recedes. In the
empty, $\omol=(0,0)$ universe, it stays at the same distance while
in the previously favored Einstein-de Sitter model, $\omol=(1,0)$,
and the $\omol=(0.3,0)$ model, it approaches. The different
behavior in each model ultimately stems from the different
compositions of the universes, because the composition dictates
the acceleration. When the cosmological constant is large enough
to cause the expansion of the universe to accelerate, the test
galaxy will also accelerate away. When the attractive force of
gravity dominates, decelerating the expansion, the test galaxy
approaches. General solutions in comoving coordinates of the
tethered galaxy problem are given by Eq.~(\ref{eq:chisolution})
and are plotted in Fig.~\ref{fig:chisolution} for the same four
models shown in Fig.~\ref{fig:Dsolution}, as well as for a
recollapsing model, $\omol=(2,0)$.

\subsection{Expansion makes the untethered galaxy join the Hubble
flow}
As shown in Fig.~\ref{fig:chisolution}, the untethered galaxy
asymptotically joins the Hubble flow in each cosmological model
that expands forever. However, Fig.~\ref{fig:Dsolution} shows that
whether the untethered galaxy joins the Hubble flow by approaching
or receding from us is a different, model dependent issue. The
untethered galaxy asymptotically joins the Hubble flow for {\em
all} cosmological models that expand forever because
\begin{equation}
\dot{D} =v_{\rm rec} + v_{\rm pec} =v_{\rm rec} + v_{\rm pec,0}/a.
\label{eq:joinflow}
\end{equation}
As $a \rightarrow \infty$ we have $\dot{D} =v_{\rm rec} = HD$,
which is pure Hubble flow. Note that the galaxy joins the Hubble
flow solely due to the expansion of the universe ($a$ increasing). 

We further see that the expansion does not effect the dynamics
because when we calculate the acceleration of the comoving galaxy,
all terms in $\dot{a}$ cancel out:
\bea
\ddot{D} &=&\dot{v}_{\rm rec} - \frac{v_{\rm
pec,0}}{a}\frac{\dot{a}}{a}\\ &=&(\ddot{a}\chi +
\dot{a}\dot{\chi}) - \dot{a}\dot{\chi} \label{eq:balance}\\
&=&\ddot{a}\chi \label{eq:onlyacceleration}\\ &=&-q\;H^2\,D,
\label{eq:q}
\eea
where the deceleration parameter $q(t)=-\ddot{a}a/\dot{a}^{2}$.
Notice that the second term in Eq.~(\ref{eq:balance}) owes its
existence to $\dot{\chi} \neq 0$ (which is only true if $v_{\rm
pec} \neq 0$) and here represents the galaxy moving to lower
comoving coordinates. The resulting reduction in recession
velocity is exactly canceled by the third term which is the decay
of the peculiar velocity. Thus all terms in $\dot{a}$ cancel, and
we conclude that the expansion, $\dot{a} > 0$, does not cause
acceleration, $\ddot{D} > 0$. Thus, the expansion does not cause
the untethered galaxy to recede (or to approach), but does result
in the untethered galaxy joining the Hubble flow ($v_{\rm
pec}\rightarrow 0$).

An alternative way to obtain Eq.~(\ref{eq:q}) is to differentiate
Hubble's Law, $\dot{D}=HD$. This method ignores $v_{\rm pec}$ and
therefore does not include the explicit cancellation of the two
terms in Eq.~(\ref{eq:balance}) of the more general calculation.
The fact that the results are the same emphasizes that the
acceleration of the test galaxy is the same as that of comoving
galaxies and there is no additional acceleration on our test
galaxy pulling it into the Hubble flow. 

\begin{figure}
\bctr \includegraphics[height=74mm,width=84mm]{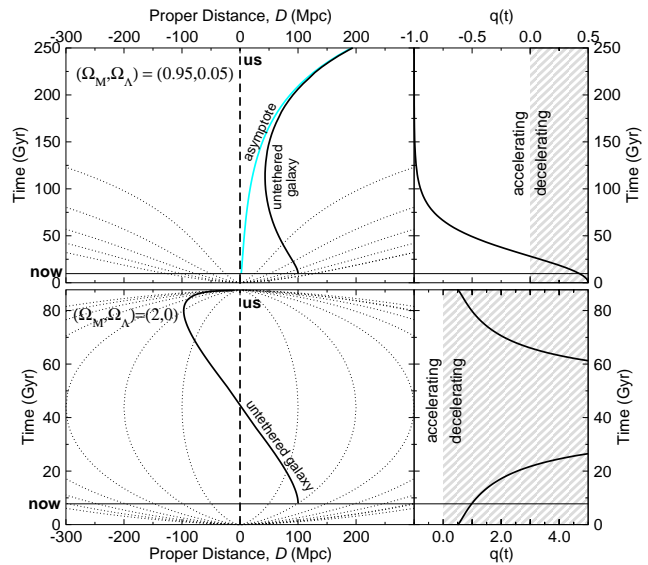}
\caption{\footnotesize{Upper panels: The deceleration parameter $q(t)$
determines the acceleration of the untethered galaxy
[Eq.~(\ref{eq:q})] and can change sign. This particular model
shows the effect of $q$ (right panel) on the position of the
untethered galaxy (left panel). Initially $q > 0$ and the proper
distance to the untethered galaxy decreases [as in an $\omol =
(1,0)$ universe], but $q$ subsequently evolves and becomes
negative, reflecting the fact that the cosmological constant
begins to dominate the dynamics of the universe. With $q < 0$, the
acceleration $\ddot{D}$ changes sign. This makes the approaching
galaxy slow down, stop, and eventually recede. The dotted lines
are fixed comoving coordinates. Lower panels: The $\omol=(2,0)$
universe expands and then recollapses ($\dot{a}$ changes sign),
and the peculiar velocity increases and approaches $c$ as
$a\rightarrow0$ [Eq.~(\ref{eq:relativistic})].}}
\label{fig:weirdexamples}\ectr
\end{figure} 
\begin{figure*}
\centering
\begin{minipage}[c]{.35\textwidth}
 \centering
 \caption{
\footnotesize{The graphs show the combination of recession velocity and peculiar velocity that result in a redshift of zero, for four cosmological models. The purpose of these graphs is to display the counter-intuitive result that in an expanding universe a redshift of zero does not correspond to zero total velocity ($\dot{D}=0$).  Gray striped areas show the surprising situations where receding galaxies appear blueshifted or approaching galaxies appear redshifted. Other models [for example, $\omol=(0.05,0.95)$, Fig.~\ref{fig:weirdexamples}, top panel] can have both approaching redshifted and receding blueshifted regions simultaneously.  Recession velocities are calculated at the time of emission; the results are qualitatively the same when recession velocities are calculated at the time of observation. Thus galaxies that were receding at emission and are still receding, can be blueshifted.  Note that in each panel for low velocities (nearby galaxies), the $z_{\rm tot} = 0$ line asymptotes to the $v_{\rm tot} = 0$ line.  See Sec.~\ref{sect:obs-cons} for a discussion of the active galactic nuclei jet data point in the upper left panel.}
 }
\label{fig:z0vpecvrec}
\end{minipage}
\hspace{3mm}
\begin{minipage}[c]{.58\textwidth}
 \centering
 \includegraphics[width=\textwidth]{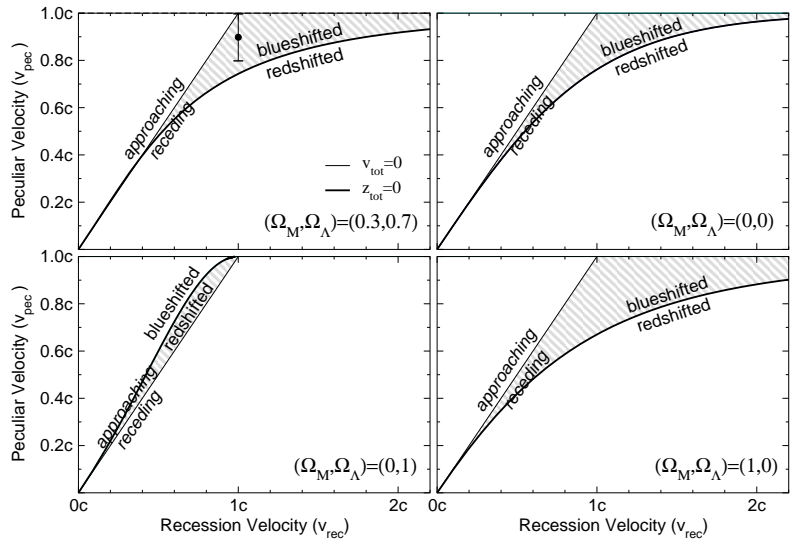}
\end{minipage}
\end{figure*}

\subsection{Acceleration of the expansion makes the untethered
galaxy approach or recede}
Because the initial condition is
$\dot{D}_0=0$, whether the galaxy approaches or recedes from us is
determined by whether it is accelerated toward us ($\ddot{D} < 0$)
or away from us ($\ddot{D} > 0$). Equation~(\ref{eq:q}) shows that
in an expanding universe, whether the galaxy approaches us or
recedes from us does not depend on the velocity of the Hubble flow
(because $H>0$) or the distance of the untethered galaxy (because
$D>0$), but on the sign of $q$. When the universe accelerates ($q
< 0$), the galaxy recedes from us. When the universe decelerates
($q > 0$), the galaxy approaches us. Finally, when $q = 0$, the
proper distance stays the same as the galaxy joins the Hubble
flow. Thus the expansion does not ``drag'' the untethered galaxy
away from us. Only the {\em acceleration of the expansion} can
result in a change in distance between us and the untethered
galaxy.

Notice that in Eq.~(\ref{eq:q}), $q= q(t) =q(a(t))$ is a function
of the scale factor:
\begin{equation}
q(a)=\biggr(\frac{\om}{2a}-\oll a^2\biggr)\biggr[1+\om\biggr(\frac{1}{a}-1\biggr)+\oll(a^2-1)\biggr]^{-1},
\end{equation}
which for $a(t_0)=1$ becomes the current deceleration parameter
$q_0 = \om/2 -\oll$. Thus, for example, the $\omol=(0.66,0.33)$
model has $q_0=0$, but $q$ decreases with time, and therefore the
untethered galaxy recedes. The upper panels of
Fig.~\ref{fig:weirdexamples} show how a changing deceleration
parameter affects the untethered galaxy. There is a time lag
between the onset of acceleration ($q < 0$) and the galaxy
beginning to recede ($v_{\rm tot} > 0$) as is usual when
accelerations and velocities are in different directions.

The example of an expanding universe in which an untethered galaxy
approaches us exposes the common fallacy that ``expanding space''
is in some sense trying to drag all pairs of points apart.  The
fact that in the $\omol=(1,0)$ universe the untethered galaxy,
initially at rest, falls through our position and joins the Hubble
flow on the other side of us does not argue against the idea of
the expansion of space.\cite{peacock99peacock01} It does, however,
highlight the common false assumption of a force or drag
associated with the expansion of space. We have shown that an
object with a peculiar velocity does rejoin the Hubble flow in
eternally expanding universes, but {\em does not feel any force}
causing it to rejoin the Hubble flow. This qualitative result
extends to all objects with a peculiar velocity.

\section{A tethered galaxy has a nonzero redshift}\label{sect:redshift}
In the context of special
relativity (Minkowski space), objects at rest with respect to an
observer have zero redshift. However, in an expanding universe
special relativistic concepts do not generally apply. ``At rest''
is defined to be ``at constant proper distance'' ($v_{\rm
tot}=\dot{D}=0$), so our untethered galaxy with $\dot{D}_0 = 0$
satisfies the condition for being at rest. Will it therefore have
zero redshift? That is, are $z_{\rm tot}=0$ and $v_{\rm tot} = 0$
equivalent? Although radial recession and peculiar velocities add
vectorially, their corresponding redshift components
combine\footnote{Light is emitted by the tethered galaxy. Let
$\lambda_{\rm observed}$ be the wavelength we observe,
$\lambda_{\rm emitted}$ be the wavelength measured in the comoving
frame of the emitter (the frame with respect to which it has a
peculiar velocity $v_{\rm pec}$) and $\lambda_{\rm rest}$ be the
wavelength of light in the rest frame of the emitter. Then
$1+z_{\rm tot} = \frac{\lambda_{\rm observed}}{\lambda_{\rm rest}}
= \frac{\lambda_{\rm observed}}{\lambda_{\rm
emitted}}\frac{\lambda_{\rm emitted}}{\lambda_{\rm rest}} =
(1+z_{\rm rec})(1+z_{\rm pec})$.} as $(1+z_{\rm tot}) = (1+z_{\rm
rec})(1+z_{\rm pec})$. The condition that $z_{\rm tot}=0$ gives
\begin{equation}
(1+z_{\rm pec}) = \frac{1}{(1+z_{\rm rec})}. \label{eq:pecrec}
\end{equation}
The special relativistic relation between peculiar velocity and
Doppler redshift is
\begin{equation}
v_{\rm pec}(z_{\rm pec}) = c \biggl[\frac{(1+z_{\rm pec})^{2} -
1}{(1+z_{\rm pec})^{2} + 1}\biggr] , \label{eq:vpeczpec} 
\end{equation}
while the general relativistic relation between recession velocity
(at emission\footnote{To calculate the current recession velocity
(as opposed to the recession velocity at the time of emission)
replace $z_{\rm rec}$ with $z=0$ in Eq.~(\ref{eq:vreczrec}) (except in the upper
limit of the integral).}) and cosmological redshift
is\cite{harrison93eq13}
\begin{equation}
 v_{\rm rec}(z_{\rm rec}) = c\; \frac{H(z_{\rm rec})}{1+z_{\rm rec}} \int_0^{z_{\rm rec}}\frac{dz}{H(z)}, \label{eq:vreczrec} 
\end{equation}
where $H(z_{\rm rec})=H(t_{\rm em})$ is Hubble's constant at the
time of emission. Hubble's constant as a function of cosmological
redshift is obtained by rearranging Friedmann's equation
[Eq.~(\ref{eq:fried})], 
\begin{equation}
H(z) = H_0(1+z)\biggr[1+\Omega_{\rm M}z +
\Omega_{\Lambda}\biggr(\frac{1}{(1+z)^2}-
1\biggr)\biggr]^{1/2}.\label{eq:Hz}
\end{equation}

In Fig.~\ref{fig:z0vpecvrec} 
we plot the $v_{\rm tot} = 0$ and the
$z_{\rm tot}$ lines to show they are not coincident. To obtain the
$z_{\rm tot} = 0$ curve, we do the following: For a given $v_{\rm
rec}$ we use Eq.~(\ref{eq:vreczrec}) to calculate $z_{\rm rec}$
(for a particular cosmological model). Equation~(\ref{eq:pecrec})
then gives us a corresponding $z_{\rm pec}$ and we can solve for
$v_{\rm pec}$ using Eq.~(\ref{eq:vpeczpec}). The result is the
combination of peculiar velocity and recession velocity required
to give a total redshift of zero.  The fact that the $z_{\rm tot}
= 0$ curves are different from the $v_{\rm tot} = 0$ line in all
models shows that $z_{\rm tot} = 0$ is not equivalent to $v_{\rm
tot} = 0$. Recession velocities due to expansion have a different
relation to the observed redshift [Eq.~(\ref{eq:vreczrec})] than
do peculiar velocities [Eq.~(\ref{eq:vpeczpec})].\cite{harrison93}

That the $z_{\rm tot} = 0$ line is not the same as the $v_{\rm
tot} = 0$ line even in the $q=0$, $\omol = (0,0)$ model (upper
right Fig.~\ref{fig:z0vpecvrec}) 
is particularly surprising
because we might expect an empty expanding FRW universe to be
well-described by special relativity in flat Minkowski spacetime.
Zero velocity approximately corresponds to zero redshift for
$v_{\rm rec} \lsim 0.3 c$ or $z_{\rm rec} \lsim 0.3$ [not just for
the $(0,0)$ model but for all models], but for larger redshifts is
not the case because of the different way time is defined in the
FRW and Minkowski metrics. A coordinate change can be made to make
the FRW model look like Minkowski spacetime, but the homogeneity
of constant time surfaces is lost.\cite{page93} As a consequence,
in the $\omol = (0,0)$ model, a galaxy at a constant distance
($\dot{D} = 0$) will be blueshifted. An analytical derivation of
the solution for the empty universe is given in
Appendix~\ref{sect:analytic-empty}.

The fact that approaching galaxies can be redshifted and receding
galaxies can be blueshifted is an interesting illustration of the
fact that cosmological redshifts are not Doppler shifts. The
expectation that when
$v_{\rm tot}=0$, $z_{\rm tot}=0$, comes from special relativity
and does not apply to galaxies in the general relativistic
description of an expanding universe, even an empty one.

\vspace{-2mm}
\section{Observational consequences}\label{sect:obs-cons}
\begin{figure}[t]
\bctr
\includegraphics[height=70mm,width=86mm]{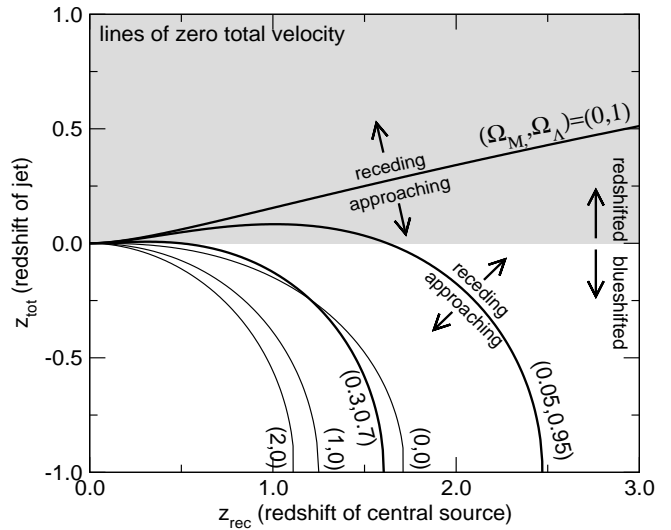}
\caption{\footnotesize{This graph expresses the same information as
Fig.~\ref{fig:z0vpecvrec}
 but in terms of observables. An active
galactic nuclei with the central source of cosmological redshift,
$z_{\rm rec}$, is assumed to be comoving. The observed redshift of
a knot in a jet, $z_{\rm tot}$, is the total redshift resulting
from the peculiar velocity of the jet and from the cosmological
redshift. The $z_{\rm tot}=0$ boundary separates the redshifted
region (upper) from the blueshifted region (lower). The curves
correspond to a total velocity of zero ($\dot{D}=0$) for different
models, $\omol$, as labeled. The regions representing receding
objects and approaching objects are indicated for the
$\omol=(0.05,0.95)$ and $\omol=(0,1)$ models as examples
(recession or approach {\it at emission} is plotted). In contrast,
for expectations based on special relativity, receding objects are
not necessarily redshifted, nor are blueshifted objects
necessarily approaching us.\vspace{-5mm}}}
\label{fig:v0ztotzrec}
\ectr
\end{figure}

The result for the tethered galaxy can be applied to the related
case of active galactic nuclei outflows. Some compact
extragalactic radio sources at high redshift are seen to have
bipolar outflows of relativistic jets of plasma. Jets directed
toward us (and in particular the occasional knots in it) are
analogs of a tethered (or boosted) galaxy. These knots have
peculiar velocities in our direction, but their recession
velocities are in the opposite direction and can be larger. Thus
the proper distance between us and the knot can be increasing.
They are receding from us (in the sense that $\dot{D} > 0$), yet,
as we have shown here, the radiation from the knot can be
blueshifted.  In Fig.~\ref{fig:v0ztotzrec} the zero-total-velocity condition is plotted in terms of the observable redshifts of a central-source and jet system. 

We can predict which radio sources have receding blueshifted jets.
The radio source 1146+531, for example, has a redshift $z_{\rm
rec}=1.629\pm0.005$.\cite{vermeulen95} In an $\omol=(0.3,0.7)$
universe, its recession velocity at the time of emission was
$v_{\rm rec} \approx c$. Therefore the relativistic jet ($v_{\rm
pec}<c$) it emits in our direction was (and is) receding from us
and yet, if the parsec scale jet has a peculiar velocity within
the typical estimated range $0.8 \lsim v_{\rm pec}/c \lsim 0.99$,
it will be blueshifted. This example is the point plotted in the
upper left panel of Fig.~\ref{fig:z0vpecvrec}.

\vspace{-1mm}
\section{Summary}
We have pointed out and interpreted some counter-intuitive results
of the general relativistic description of our Universe. We have
shown that the unaccelerated expansion of the universe has no
effect on whether an untethered galaxy approaches or recedes from
us. In a decelerating universe the galaxy approaches us, while in
an accelerating universe the galaxy recedes from us. The
expansion, however, {\em is} responsible for the galaxy joining
the Hubble flow, and we have shown that this happens whether the
untethered galaxy approaches or recedes from us.

The expansion of the universe is a natural feature of general
relativity that also allows us to unambiguously convert observed
redshifts into proper distances and recession velocities and to
unambiguously define approach and recede. We have used this
foundation to predict the existence of receding blueshifted and
approaching redshifted objects in the universe. To our knowledge
this is the first explicit derivation of this counter-intuitive
behavior.

Concepts such as ``recede'' or ``approach'' and quantities such as
$\dot{D}$ are of limited use in observational cosmology because
all our observations come to us via the backward pointing null
cone. This limitation will remain the case until a very patient
observer organizes a synchronized set of comoving observers to
measure proper distance.\cite{weinberg72rindler77} However, the
issue we are addressing --- the relationship between observed
redshifts and expansion --- is a conceptual one and is closely
related to the important conceptual distinction between the
theoretical and empirical Hubble laws.\cite{harrison93}
\newline\newline
{\bf Postscript:}\newline
Post publication it was brought to our attention that T. Kiang 
has published a similar analysis of the redshifts of relativistic jets.  
His excellent paper can be found in Chinese Astronomy and Astrophysics 
25, 141--146 (2001).
\vspace{-6mm}
\begin{acknowledgments}\vspace{-5mm}
TMD is supported by an Australian Postgraduate Award. CHL
acknowledges an Australian Research Council Fellowship. JKW
acknowledges useful discussions with Ken Lanzetta. We wish to
thank Jochen Liske, John Peacock, Phillip Helbig, Edward Harrison,
 Joe Wolfe, and Frank Tipler for useful discussions and two anonymous referees for constructive comments. \end{acknowledgments}

\appendix \section{Luminosity Distance, Angular-Diameter
Distance}\label{ap:distances}
The FRW metric including the angular
terms is
\begin{equation}
ds^2=-c^2dt^2+a^2(t)[d\chi^{2} + S^{2}_{k}(\chi) (d\theta^{2} +
\chi^{2}d\phi^{2})] ,
\end{equation}
where $S_{k}(\chi) = \sinh \chi$, $\chi$, $\sin \chi$ for $k= -1$,
0, 1, respectively, and $\theta$ and $\phi$ are the angular
measures in spherical coordinates. We use the proper distance
$D=a\chi$, which is the distance measured along a spatial
geodesic, the path light follows through space. Other distance
measures in common use are angular diameter distance $D_{A} =
(1+z)^{-1} a(t)S_{k}(\chi)$ and luminosity distance $D_{L} = (1+z)
a(t)S_{k}(\chi)$. Both include the $S_{k}$ term, which means they
both involve the distance perpendicular to the line of sight.
[$S_{k}(\chi)$ appears only in the metric when multiplied by an
angular term.] They can be used to parametrize distance, but have
no direct relation to recession velocity and cannot be used to
explain the observed redshift. The distance in Hubble's law,
$v_{\rm rec} = HD$, is proper distance. If one prefers to use
$D_{A}$ or $D_{L}$ as measures of distance and $\dot{D}_{L}$ and
$\dot{D}_{A}$ to define ``approach'' and ``recede,'' it can also
be shown, using the relationships between $\dot{D}$,
$\dot{D}_{A}$, and $\dot{D}_{L}$, in a fashion similar to what we
have done for proper distance, that $z_{\rm tot} = 0$ is not
equivalent to either $\dot{D}_{L} = 0$ or $\dot{D}_{A} = 0$.

\section{RELATIVISTIC SOLUTION FOR PECULIAR
VELOCITY}\label{ap:rel-sol}
When a universe collapses, the scale
factor $a$ decreases. Thus
$v_{\rm pec}\propto 1/a$ [see Eq.~(\ref{eq:Rinv})] means that the
peculiar velocity increases with time. Therefore, in collapsing
universes, untethered galaxies do not ``join the Hubble flow.''
This behavior is shown for the $\omol=(2,0)$ model in
Fig.~\ref{fig:chisolution}. Collapsing universes require the
relativistic formula for the change of the peculiar velocity to
avoid the infinite peculiar velocities that result from $v_{\rm
pec}\propto 1/a$ as $a\rightarrow0$. To produce all the figures in
this paper, except the lower panels in Fig.~\ref{fig:weirdexamples}, we have used
$p=mv_{\rm pec}$. However, as the peculiar velocities become
relativistic in a collapsing universe, we need to use the special
relativistic formula for momentum $p=\gamma mv_{\rm pec}$, where
$\gamma=(1-v_{\rm pec}^2/c^2)^{-1/2}$. Because momentum decays as
$1/a$ ($p=p_{\rm o}a_0/a$), we obtain, 
\begin{equation}
v_{\rm pec}= \frac{\gamma_0 v_{\rm pec,0}}{\sqrt{a^2+\gamma_0^2
v_{\rm pec_0}^2/c^2}} . \label{eq:relativistic}
\end{equation}
Therefore, as $a\rightarrow0$, $v_{\rm pec}\rightarrow c$.
Equation~(\ref{eq:relativistic}) was used to produce the lower
panels of Fig.~\ref{fig:weirdexamples}.
The relativistic formula for momentum should also be used in eternally expanding universes if relativistic velocities are set as the initial condition in Eq.~(\ref{eq:initial1}).
Using Eq.~(\ref{eq:relativistic}) in Eq.~(\ref{eq:joinflow}) results in a residual dependence on $\dot{a}$ in Eq.~(\ref{eq:onlyacceleration}).  The residual is negligible for $v<\hspace{-1.5mm}<c$, and becomes negligible for $v\sim c$ as $a\rightarrow \infty$.   
Note that Eq.~(\ref{eq:vpeczpec}) is relativistic and therefore the results of Section~\ref{sect:redshift} hold for $v_{\rm pec}\sim c$.

Collapsing universes also provide the possibility of
approaching-redshifted objects, but without involving peculiar
velocities. In the collapsing phase all galaxies are approaching
us. However, if the galaxy is distant enough, it may have been
receding for the majority of the time its light took to propagate
to us. In this case the galaxy appears redshifted even though it
may be approaching at the time of observation. This example
differs from the active galactic nuclei jet example because the
active galactic nuclei jet may appear blueshifted even though the
jet {\em never approaches us}.

\vspace{-2mm}
\section{ANALYTIC SOLUTION FOR THE EMPTY
UNIVERSE}\label{sect:analytic-empty}
In the empty $\omol=(0,0)$
universe, an analytical solution can be found for the combination
of recession and peculiar velocity that would give a redshift of
zero. For an empty expanding universe,
$H(z)=H_0(1+z)$, and the time derivative of the scale factor at
emission is $\dot{a}_{\rm em}=H_0$. Therefore,
Eq.~(\ref{eq:vreczrec}) becomes
\bea
v_{\rm rec}&=&c H_0 \! \int_0^{z_{\rm rec}}\frac{dz}{H_0(1+z)} \\
&=&c \ln(1+z_{\rm rec}) \\ e^{v_{\rm rec}/c}&=& 1+z_{\rm
rec}.\label{eq:evrec}
\eea
If we substitute Eq.~(\ref{eq:evrec}) into Eq.~(\ref{eq:pecrec})
followed by Eq.~(\ref{eq:vpeczpec}), we find
\begin{equation}
v_{\rm pec}=c\left[\frac{e^{-2v_{\rm rec}/c}-1}{e^{-2v_{\rm
rec}/c}+1}\right].\label{eq:vpec00} 
\end{equation}
Equation~(\ref{eq:vpec00}) shows that only in the limit of small
$v_{\rm rec}$ does $v_{\rm pec}= -v_{\rm rec}$ for $z_{\rm
tot}=0$. Equation~(\ref{eq:vpec00}) generates the thick black $z=0$
line in Fig.~\ref{fig:z0vpecvrec}, 
upper right panel.

\vspace{-2mm}
\section{SUGGESTED PROBLEMS}
The host galaxy of active galactic nuclei 1146+531 has a
redshift $z_{\rm rec} = 1.63$. Assume for simplicity that we live
in a universe with $\omol=(1,0)$.

(a) What was the galaxy's recession velocity at the time it
emitted the light we now see? [Refer to Eqs.~(\ref{eq:vreczrec})
and (\ref{eq:Hz}).]

(b) If the jet it emits had a peculiar velocity in our direction of
$v_{\rm pec} = 0.80c$, what was the jet's total velocity at the
time of emission? [Refer to Eq.~(\ref{eq:vrecvpec}).] Is it moving
away from or toward us?

(c) What is the jet's total redshift, $z_{\rm tot}$? [See
Eq.~(\ref{eq:vpeczpec}) and the text preceding
Eq.~(\ref{eq:pecrec}).] Is it redshifted or blueshifted?

(d) What is the galaxy's recession velocity at the time of
observation?$^{24}$ Compared to your answer in Part (a) is this
behavior what you would expect for a decelerating universe?

\newpage

\end{document}